\documentclass[a4paper]{article}

\usepackage{INTERSPEECH2018}
\usepackage{multicol}
\usepackage{multirow}
\usepackage{color}
\usepackage{xcolor}
\usepackage{soul}
\usepackage{times}
\usepackage{rotating}

\usepackage{graphicx}
\usepackage[normalem]{ulem}

\title{Classification of Huntington Disease using Acoustic and Lexical Features}

\name{Matthew Perez $^1$, Wenyu Jin $^1$, Duc Le $^1$, Noelle Carlozzi $^2$, Praveen Dayalu $^3$, Angela Roberts $^4$, Emily Mower Provost $^1$
}
\address{Computer Science and Engineering, University of Michigan, Ann Arbor, MI $^1$\\Physical Medicine \& Rehabilitation, University of Michigan, Ann Arbor, MI $^2$ \\Michigan Medicine, University of Michigan, Ann Arbor, MI $^3$ \\Department of Communication Sciences and Disorders, Northwestern University, Evanston, IL$^4$ }

\email{mkperez@umich.edu, wyjin@umich.edu, ducle@umich.edu, carlozzi@med.umich.edu, pravd@med.umich.edu, angela.roberts@northwestern.edu emilykmp@umich.edu}

\begin{document}

\maketitle
\begin{abstract}
Speech is a critical biomarker for Huntington Disease (HD), with changes in speech increasing in severity as the disease progresses.  Speech analyses are currently conducted using either transcriptions created manually by trained professionals or using global rating scales.  Manual transcription is both expensive and time-consuming and global rating scales may lack sufficient sensitivity and fidelity~\cite{carlozzi2016hdqlife}. Ultimately, what is needed is an unobtrusive measure that can cheaply and continuously track disease progression.  We present first steps towards the development of such a system, demonstrating the ability to automatically differentiate between healthy controls and individuals with HD using speech cues.  The results provide evidence that objective analyses can be used to support clinical diagnoses, moving towards the tracking of symptomatology outside of laboratory and clinical environments.
\end{abstract}


\vspace{-3pt}
\section{Introduction}
\vspace{-3pt}
\label{sec:intro}
Huntington disease (HD) is a fatal, autosomal neurodegenerative disease that typically affects 12 per 100,000 people in the Western World~\cite{mason2018predicting}. HD is insidious and progressive, affecting motor skills, speech, cognition, and behavior~\cite{paulsen2010early}. The diagnosis of HD is based on unequivocal motor symptoms and is typically made when individuals are in their mid-40's~\cite{hinzen2017systematic,vogel2012speech}.  
Current research suggests that speech motor deficits precede the onset of limb and trunk chorea~\cite{vogel2012speech}, providing an opportunity to leverage changes in speech as a sensitive biomarker.  These biomarkers can then be used to support distributed ecologically valid symptom tracking. This paper builds towards that goal, investigating how the speech signal can be used to automatically detect HD.


HD speech is typically characterized by decreases in the number of words pronounced, syntactic complexity, and speech rate, in addition to increases in paraphasic errors, filler usage, and sentence duration~\cite{HD_speech_characterization, vogel2012speech}. Language and speech symptoms are common in HD, occurring in approximately 90\% of cases~\cite{dysarthria_1, dysarthria_2}. As such, acoustic analyses may provide meaningful therapeutic and diagnostic information for individuals with HD, especially given that preliminary research has shown that HD-related speech deficits can be objectively characterized and increase with disease progression~\cite{kaploun2011acoustic}.  The development of an objective, non-invasive acoustic biomarker, sensitive enough to detect disease progression in people with premanifest and manifest HD, will provide new avenues for clinical research and treatments. 

We present an initial step towards detecting changes in HD severity by first demonstrating the efficacy of the speech signal for detecting the presence of HD.  The system includes transcription, feature extraction, and classification. The transcripts are generated either by humans or by training in-domain automatic speech recognition (ASR) systems.  The features are clinically-inspired and include filler usage, pauses in speech, speech rate, and pronunciation errors. We investigate the static and dynamic feature properties.  The static feature sets describe speech behavior using summary statistics; the dynamic feature sets provide an opportunity to directly model the feature variation between utterances. We model the static feature sets using k-Nearest Neighbors ($k$-NN) with Euclidean distance and Deep Neural Networks (DNN). We model the dynamic feature sets using $k$-NN with Dynamic Time Warping (DTW) Distance and Long-Short-Term Memory Networks (LSTM).  We investigate the impact of transcription error, moving from manual transcriptions, to transcripts generated using forced alignment to known prompts, and finally to automatic speech recognition (ASR).

Our results demonstrate the efficacy of speech-centered approaches for detecting HD.  We show that we can accurately detect HD using a simple static ($k$-NN) approach, resulting in an accuracy of 0.81 (chance performance is 0.5).  We then show that these results can be improved using dynamic feature sets (DTW) or deep methods, which can capture the non-linear relationships in our feature sets (DNN/LSTM).  Finally, we demonstrate that in domains with limited lexical variability, manual transcripts can be replaced with ASR transcripts without suffering from a degradation in performance, even given a word error rate of 9.4\%.  This indicates the robustness of the identified speech features. The novel aspects of our approach include one of the first investigations into automated speech-centered HD detection and a focus on understanding the importance of modeling temporal variability for detecting HD.

\begin{table*}[t]
\vspace{-5pt}
\caption{Summary of participant demographics}
\vspace{-5pt}
\label{tab:speaker_data}
\centering
\begin{tabular}{c | c c c c c c}
\hline
 & Premanifest (n=12) & Early (n=12) & Late (n=7) & Control (n=31) & All (n=62) \\
\hline
\hline
Age | Mean (SD) & 42.6 (9.8) & 52.0 (11.2) & 54.6 (9.7) & 50.3 (11.1) & 49.8 (11.1) \\
Gender | \% Male & 36.4 & 41.7 & 37.5 & 38.7 & 38.7 \\
Race | \% White & 90.9 & 100 & 87.5 & 90.3 & 91.9 \\
Race | \% Black & 0 & 0 & 12.5 & 6.5 & 4.8 \\
Race | \% Other & 9.1 & 0 & 0 & 3.2 & 3.3 \\
\hline
\end{tabular}
\vspace{-6pt}
\end{table*}

\vspace{-3pt}
\section{Related Work}
\vspace{-3pt}
\label{sec:rw}

\begin{table}[t]
\vspace{-3pt}
\caption{Dataset summary: total dataset size (seconds) and average utterance duration (seconds). HC describes the healthy controls and HD, the participants with Huntington}
\vspace{-3pt}
\label{tab:utt_data}
\centering
\begin{tabular}{c | c c }
\hline
Group & Duration (avg. dur) & \# Utt. (avg. num) \\
\hline
\hline
HC & 1,469.83 (47.41 $\pm$ 7.68) & 310 (10.0 $\pm$ 0.00)  \\
HD & 2,641.83 (85.2 $\pm$ 38.18) & 320 (10.32 $\pm$ 0.74)  \\
\hline
\end{tabular}
\vspace{-16pt}
\end{table}


Previous works have demonstrated the feasibility of automated speech assessment for various neurocognitive disorders such as dementia~\cite{dementia}, aphasia~\cite{aphasia}, and Alzheimer's disease~\cite{toth2015automatic}. Research has investigated the feasibility of using automatic speech recognition (ASR) to extract lexical features from  transcribed audio~\cite{fraser2013automatic, sadeghian2015using}. However, off-the-shelf ASR systems are not well suited to this domain because of the abnormal speech patterns, high speaker variability, and lack of data that are common in dysarthric speech~\cite{zhou2016speech,mengistu2011comparing}. Furthermore, these off-the-shelf ASR systems may miss out on critical cues such as the presence of fillers, stutters, or mispronunciations, all of which contribute to the perception of disordered speech~\cite{toth2015automatic}. Another approach has been to train specialty ASR models directly on the domain-relevant data.  Le et al. trained an ASR system on aphasic speech and used this system to extract speech features for the prediction of intelligibility, including: the number of fillers, phone rate, error rate, utterance length, and goodness of pronunciation~\cite{aphasia}. Similar work from Peintner et al. focused on the automatic diagnosis of neurodegenerative diseases~\cite{peintner2008learning}. In addition, work from Guerra et al. demonstrated the potential of using non-linear classifiers and multi-dimensional features to detect dysarthric speech.~\cite{guerra2003modern}.

\vspace{-3pt}
\section{Data Description}
\vspace{-3pt}
\label{sec:data}
The data in this study was collected from an HD study conducted at the University of Michigan. The data consists of 62 speakers, 31 healthy and 31 with HD. Out of the 31 individuals with HD, 11 are premanifest, 12 are in the early stage, and 8 are in the late stage. HD groups were created using the Total Motor Score (TMS) and the Total Functional Capacity (TFC) score \cite{shoulson1989assessment} from the Unified Huntington’s Disease Rating Scale (UHDRS) \cite{kieburtz2001unified}. Specifically, individuals were designated as premanifest HD if they had a positive gene test (HD CAG $>$ 35) and a clinician-rated score of less than 4 on the last item of the TMS (which provides an index of clinician-rated diagnostic confidence).  Those with clinician-rated scores greater than or equal to 4 on the last item of the TMS were included in the manifest HD group.  For those with manifest HD, TFC scores (which provide an index of clinician-rated functional capacity) were used to determine HD stage. Specifically, scores range from 0 (low functioning) to 13 (highest level of functioning); TFC sum scores of 7-13 were considered early-stage and  sum scores of 0-6 were considered late-stage HD.


The data includes both read speech and spontaneous speech sections.  This study focuses only on the read speech portion, during which participants read the Grandfather Passage~\cite{Grandfather_p1}. The Grandfather Passage~\cite{duffy1995motor, zraick2004reliability, darley1975motor} is a phonetically balanced paragraph containing 129 words and 169 syllables, and is a standard reading passage used in speech-language pathology. This passage is commonly used to test for dysarthric speech~\cite{patel2013caterpillar}.  Table~\ref{tab:utt_data} shows additional information about the scope of our data such as the size and number of utterances.

\section{Data Transcription}
\label{sec:trans}

\subsection{Human Transcriptions}
\label{sec:orat}

Recordings were deidentified and transcribed using the CHAT approach~\cite{macwhinney2000childes} and Computerized Language Analysis (CLAN) software. CHAT transcriptions identify speech errors (phonological, semantic, or neologistic), vowel distortions, word repetitions, retracing, assimilations, dialect variances, letter and word omissions, utterances, pauses, glottal sounds unique to HD, vocalizations, spontaneous speech for each participant, and variations in rate, fundamental frequency (F0), and voice quality. Interrater reliability (greater than or equal to 90\% agreement) was established between two trained raters and a Ph.D.-level Speech Language Pathologist. The raters then individually transcribed each recording and their transcriptions were compared. Raters were required to reach a consensus for all identified discrepancies. In cases where consensus could not be reached, the Speech Language Pathologist was consulted.

\begin{table}[t]
\vspace{-3pt}
\caption{Performance of ASR system (per speaker) for healthy and HD speech, including WER, insertion (ins), deletion (del), and substitution (sub)}
\vspace{-3pt}
\label{tab:wer}
\centering
\begin{tabular}{c | c c c c}
\hline
 &WER \%&Ins&Del&Sub\\
\hline
All  & $9.4\pm14.8$ & $1.4\pm3.1$ & $5.0\pm11.2$ & $8.9\pm17.3$\\
HD & $16.0\pm18.7$ & $2.5\pm4.0$ & $8.7\pm14.9$ & $15.5\pm22.5$\\
HC & $2.8\pm2.3$ & $0.3\pm0.9$ & $1.3\pm1.2$ & $2.3\pm2.4$\\
\hline
\end{tabular}
\vspace{-16pt}
\end{table}

\begin{figure*}[t]
\begin{center}
\includegraphics[scale=0.63]{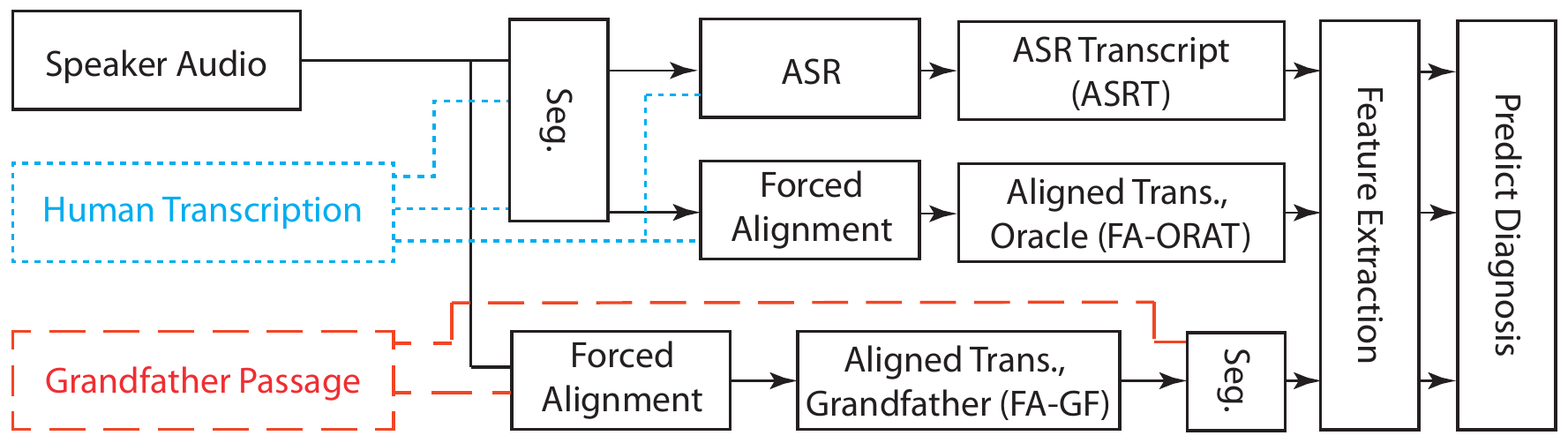}
\vspace{-3pt}
\caption{System diagram for HD classification (note: seg. is segmentation) \label{fig:data_pipeline}}
\end{center}
\vspace{-16pt}
\end{figure*}

\vspace{-3pt}
\subsection{Automated Transcripts}
\vspace{-3pt}
\label{sec:asrt}

Manual transcripts often represent a bottleneck because they are costly and time-intensive to obtain. ASR can provide an alternative. However, off-the-shelf systems are often unusable due to the acoustic mismatch between the healthy speech used to train these systems and the speech patterns of individuals of the target population. We address this by training in-domain acoustic and language models using a specialized lexicon. 

The lexicon we used is initialized using standard English phone-level pronunciations provided by the CMU pronunciation dictionary \cite{weide1998cmu}.  We augment this using the pronunciation errors identified in the manual transcripts.  We use a bigram language model extracted over the manual transcripts.

The acoustic model is a monophone Hidden Markov Model (HMM) with three-states per phone and a left-to-right topology. The emission probability of each state is estimated using a Gaussian Mixture Model (GMM).  We use a monophone acoustic model rather than a triphone model due to the relatively small size  of the dataset ($\sim$1 hour of total speech). The final model consists of 500 Gaussian mixtures.

The input acoustic features are Mel-frequency Cepstral Coefficients (MFCC), extracted with a 25ms Hamming window and 10ms overlap using Kaldi~\cite{povey2011kaldi}. We include the 13 MFCC coefficients, the first, second, and third order deltas.

The acoustic and language models are trained and tested using a Leave-One-Subject-Out (LOSO) approach over all HC and HD speakers.  The performance of our ASR system is shown in Table~\ref{tab:wer}, obtaining an overall WER of $9.4\pm14.9\%$. This relatively low WER is strongly attributed to the constrained speech produced by reading the grandfather passage.

\vspace{-3pt}
\subsection{Preprocessing Methods}
\vspace{-3pt}

We present three approaches to investigate the effect of error propagation.  We first assume the availability of manual transcripts and investigate the feasibility of extracting features for HD classification (\emph{force-aligned oracle transcription}, FA-ORAT).  Next, we assume that the subject prompt is available and ask if the subject's speech goals can be used as a target for force alignment (\emph{forced-aligned grandfather transcription}, FA-GF). This allows us to investigate the effect of the mismatch introduced by speech errors.  Finally, we assume only that segmentation information is available, noting when speaker utterances begin and end, but that the transcript itself is unknown (\emph{ASR transcription}, ASRT). This provides insight into the effectiveness of an automatic system that can transcribe, extract features, and predict diagnosis (Figure~\ref{fig:data_pipeline}).

All approaches use the acoustic model discussed in Section~\ref{sec:asrt}. The ASRTs are discussed in Section~\ref{sec:asrt}.  FA-ORATs are generated by force-aligning the input audio to the manual transcripts.  FA-GFs are generated by force-aligning the input audio to the original Grandfather passage text.  

FA-ORAT and ASRT utterances are segmented using the manual transcripts before both ASR and force alignment (future work will remove the reliance on manual segmentation for ASRT).  FA-GF utterances are segmented into utterances using the natural sentences within the Grandfather passage ( Figure~\ref{fig:data_pipeline}). 


\begin{table}[t]
\vspace{-2pt}
\caption{Average dimension size of feature vectors}
\vspace{-3pt}
\label{tab:dim-size}
\centering
\begin{tabular}{c | c c c}
\hline
 &FA-ORAT & FA-GF & ASRT\\
\hline
Utt-level  & $43.7\pm1.2$ & $29.2\pm1.3$ & $44.3\pm1.1$\\
Spkr-level & $336.8\pm10.6$ & $235.9\pm8.0$ & $359.0\pm8.0$\\
\hline
\end{tabular}
\vspace{-10pt}
\end{table}

\vspace{-3pt}
\section{Feature Extraction}
\vspace{-3pt}
\label{sec:feat}

We describe two feature sets: utterance-level (dynamic) and speaker-level (static). The utterance-level features are extracted over each utterance and provide insight into the relationship between the time-series behavior of the features and diagnosis. We normalize the utterance-level features using speaker-dependent z-normalization.  The speaker-level features are calculated by applying summary statistics to the normalized utterance-level features, including max, min, mean, SD, range, and quartiles (25th, 50th, 75th). We group all features by subject and remove features that have either zero variance or zero information gain with respect to the target class (Table~\ref{tab:dim-size}). We perform this once at the utterance-level for our dynamic features and again after summary statistics are computed for our static features.
\vspace{4pt}

\noindent\textbf{Filler Features}: Fillers are parts of speech that are not purposeful and that do not contain formal meaning (i.e., ah, eh, um, uh). Fillers are labeled during the human transcription process.  They are preserved in ORAT and are estimated in ASRT. They are ignored in FA-GF due to the absence of fillers in the original passage. The utterance-level features include: number of fillers, number of fillers per second, number of fillers per word, number of fillers per phone, total filler duration per utterance, and total filler duration per second. 

\vspace{4pt}
\noindent\textbf{Pause Features}: Pauses are periods without speech that last for at least 150 ms~\cite{roark2011spoken}.  The utterance-level features include: the number of pauses, number of pauses per second, number of pauses per word, number of pauses per phone, total pause duration, and total pause duration per second. 

\vspace{4pt}
\noindent\textbf{Speech Rate Features}: Speech rate captures an individual's speaking speed. The utterance-level features include: the number of phones, number of phones per second, number of phones per word, number of words, and number of words per second. 

\vspace{4pt}
\noindent\textbf{Goodness of Pronunciation Features}: Goodness of Pronunciation (GoP) measures the fitness of a reference acoustic model (trained over all HD and HC speakers) to a given phone by computing the difference between the average acoustic log-likelihood of a force-aligned phoneme and that of an unconstrained phone loop~\cite{GoP}: 
\vspace{-2pt}

\begin{equation}
\vspace{-1pt}
\text{GoP}(\mathbf{p}) = \frac{1}{N}\text{log}\frac{P(O|\mathbf{p})}{P(O|PL)},
\label{eq:GoP}
\end{equation}

\noindent where $\mathbf{p}$ is a sequence of phones, $O$ is the MFCC acoustic observation, $N$ is the number of frames, and $PL$ is the unconstrained phone loop. The utterance-level features include: the GoP score for each phone in the utterance.  



\begin{table*}
\vspace{-2pt}
\caption{Classification results (FA = forced alignment, ORAT = oracle, GF = grandfather, ASRT = ASR system)}
\label{tab:results-new}
\centering
\begin{tabular}{ c | c | c | c |c|c|c} \hline
\multirow{2}{*}{Method}&
\multicolumn{2}{c|}{FA-ORAT}&
\multicolumn{2}{c|}{FA-GF}&
\multicolumn{2}{c}{ASRT}\\
\cline{2-7}
     & Accuracy & F1 (HD) & Accuracy & F1 (HD) &
     Accuracy & F1 (HD) \\ 
     \hline
     \hline
    $k$-NN & 0.81 & 0.77 & 0.82 & 0.79 & 0.81 & 0.77 \\ 
    DTW & 0.87 & 0.86 & 0.84 & 0.81 & 0.81 & 0.77\\ 
    DNN & 0.87 & 0.87 & 0.85 & 0.84 &  0.85 & 0.84\\ 
    LSTM-RNN & 0.87 & 0.86 & 0.84 & 0.82 & 0.85 & 0.84 \\
    \hline
    \hline
\end{tabular}
\vspace{-10pt}
\end{table*}

\vspace{-3pt}
\section{Methods}
\vspace{-3pt}
\label{sec:methods}

We use Leave-One-Subject-Out (LOSO) paradigm: in each run, a single subject is held-out as the test speaker and the model is trained and validated on the remaining speakers.  Within the training partition, 80\% of the data is used to train the model and 20\% is used to validate.  This process is repeated over all speakers.  All results presented are accuracies averaged over all subjects in the study.  We train four models: $k$-NN with Euclidean distance ($k$-NN), $k$-NN with DTW distance (DTW), Deep Neural Networks (DNN), and Long-Short-Term Memory Recurrent Neural Networks (LSTM-RNN).  

We hypothesize that HD can be detected using speech features.  We test this hypothesis first using $k$-NN, which assigns a label to an instance based on the plurality of its closest $k$ neighbors.  $k$-NN uses the speaker-level features, while DTW uses the utterance-level features.  In both approaches, we sweep over the number of neighbors, $k$.

We further hypothesize that the relationship between speech features and diagnosis can be more accurately modeled by exploiting non-linear feature interactions.  We test this hypothesis using Deep Neural Networks (DNN) over the speaker-level features and Long-Short-Term Memory Recurrent Neural Networks (LSTM-RNN) over the utterance-level features.  Both DNN and LSTM-RNN are implemented using Keras with a Tensorflow~\cite{abadi2016tensorflow} backend. The DNN is comprised of two fully connected layers with ReLU activation functions, a softmax output layer for binary classification, and dropout layers between each fully connected layer in the network. Our LSTM-RNN is comprised of a two LSTM layers with recurrent dropout, bias l2 regularization, and kernel l2 regularization followed by a softmax output layer.  In both networks, we perform a hyperparameter sweep for layer width (32, 64, 128) and dropout rate (0.0, 0.2, 0.4). We use an ensemble approach, in which we train five separate models and the mode of the five predictions is used as the final prediction of the system.

\begin{table}[t]
\vspace{-2pt}
\caption{The confusion matrix derived from the average percentage of classifications across all data processing methods and classifiers. Rows = ground truth, columns = prediction}
\label{tab:conf}
\centering
\begin{tabular}{c | c | c }
\hline
& Healthy & HD \\ 
\hline
\hline
Healthy & 0.95 & 0.05  \\ 
Premanifest & 0.54  & 0.46  \\ 
Early & 0.14  & 0.86  \\ 
Late & 0.02 & 0.98 \\ 
\hline
\end{tabular}
\vspace{-14pt}
\end{table}

\vspace{-3pt}
\section{Results}
\label{sec:results}
\vspace{-3pt}

The results (Table~\ref{tab:results-new}) show the feasibility of HD detection using all presented methods. 
The FA-ORAT approach results in the most accurate HD predictions, with an accuracy of 0.81 for $k$-NN, 0.87 for DTW, 0.87 for DNN, and 0.87 for LSTM-RNN. In the majority of cases in Table~\ref{tab:results-new}, the accuracy slightly decreases when less accurate transcriptions are used in the place of ORAT.  For example, we obtain an accuracy of 0.85 for DNN when using both FA-GF and ASRT. 


We assess the statistical significance of the changes in performance across classification and transcription approaches using Cochran's Q test, which compares the binary prediction over each of the 62 speakers across all methods and classifiers.  We assert significance when $p<0.05$. The result demonstrates that there is not an overall statistically significant difference over individual classifiers and transcription method (Q(11)=15.6, p=0.157). This suggests that there are multiple opportunities to recognize symptomatology and avenues to research how speech changes are associated with illness. Further, given appropriately constrained content, ASR transcripts can be used as a substitute to manual transcripts for extracting speech features to assess HD symptomatology.


\begin{table}[t]
\vspace{-2pt}
\caption{Percentage of features (GoP, Speech Rate (SR), Pauses (P)) extracted from ORAT that are statistically significantly different (p$<$0.05) compared to the HC population}
\label{tab:features}
\centering
\begin{tabular}{c | c| c |c |  c } \hline
\multirow{2}{*}{Feature}&\multicolumn{1}{c|}{}&\multicolumn{3}{c}{ORAT}\\
\cline{2-5}
 & Feature & Increase & Decrease & No Change \\
     \hline\hline
    \multirow{3}{*}{\begin{turn}{90}Pre\end{turn}} & GoP & 0.0 & 0.06 & 0.94 \\ 
    & SR & 0.04 & 0.32 & 0.64\\ 
    & P & 0.35 & 0.0 & 0.65 \\ 
    \hline
    \multirow{3}{*}{\begin{turn}{90}Early\end{turn}} & GoP & 0.12 & 0.76 & 0.12 \\ 
    & SR & 0.2 & 0.48 & 0.32 \\
    & P & 0.85 & 0.0 & 0.15\\ 
    \hline
    \multirow{3}{*}{\begin{turn}{90}Late\end{turn}} & GoP & 0.12 & 0.76 & 0.12 \\ 
    & SR & 0.2 & 0.6 & 0.2 \\ 
    & P & 0.64 & 0.0 & 0.36 \\ 
    \hline
\end{tabular}
\vspace{-14pt}
\end{table}




 
Our system is accurately able to distinguish between healthy patients and individuals with early and late stage HD (Table~\ref{tab:conf}). Our results show improved classification for later HD stages, which suggests that our features can more accurately capture HD speech for individuals whose disease is more advanced, compared to those at earlier stages.  Further, it points to the difficulty in recognizing premanifest HD due to similarities in speech compared to both healthy and HD populations.

We analyze the relationship between feature category and disease stage, focusing on the static ORAT feature set.  We aggregate the test sets generated over each run of LOSO (62 sets), retaining only the features that have non-zero variance and information gain across all 62 speakers (GoP, speech rate, pause, and filler features).  We then separate the data into the four disease categories (HC, premanifest, early, late) and identify the subset of features that are significantly different between the HC population and each of the disease stages.  We assert significance when $p<0.05$, using a two-tailed independent samples t-test. We apply Bonferroni correction to account for the family-wise error rate. We present the percentage of features that are statistically significant between the HC population and each disease stage and note whether the features of the individuals with HD are greater than, less than, or not statistically significantly different from those of the HC population. The results demonstrate that generally, GoP decreases, speech rate decreases, and the number of pauses increase with disease severity (Table~\ref{tab:features}).

\vspace{-3pt}
\section{Conclusion}
\vspace{-3pt}
\label{sec:conc}

In this work, we demonstrate the effectiveness of classifying HD using key speech features, including speech rate, pauses, fillers, and GoP.  Our experimental results show that automated approaches can be used to generate transcripts, extract features, and classify HD.  The accuracy of the presented method increases with disease stage, which suggests that speech may serve as an effective biomarker that could be used to track HD progression. Finally, the performance of both the static and dynamic approaches suggests that there are mulitple opportunities for tracking symptomatology in this domain.  Further improvements for our automated system can be made by increasing the performance of ASR by incorporating additional out-of-domain data (e.g.,~\cite{le2017automatic}). We will also investigate the development of new, more descriptive, features.

\vspace{-3pt}
\section{Acknowledgements}
\vspace{-3pt}
Work on this manuscript was supported by the National Institutes of Health (NIH), National Center for Advancing Translational Sciences (UL1TR000433). In addition, a portion of this study sample was collected in conjunction National Institutes of Health (NIH), National Institute of Neurological Disorders and Stroke (R01NS077946) and/or Enroll-HD (funded by the CHDI Foundation). Lastly, this work was also supported by the National Science Foundation (CAREER-1651740). 

\bibliographystyle{IEEEtran}
\bibliography{HD}

\begin{thebibliography}{10}
\providecommand{\url}[1]{#1}
\csname url@samestyle\endcsname
\providecommand{\newblock}{\relax}
\providecommand{\bibinfo}[2]{#2}
\providecommand{\BIBentrySTDinterwordspacing}{\spaceskip=0pt\relax}
\providecommand{\BIBentryALTinterwordstretchfactor}{4}
\providecommand{\BIBentryALTinterwordspacing}{\spaceskip=\fontdimen2\font plus
\BIBentryALTinterwordstretchfactor\fontdimen3\font minus
  \fontdimen4\font\relax}
\providecommand{\BIBforeignlanguage}[2]{{%
\expandafter\ifx\csname l@#1\endcsname\relax
\typeout{** WARNING: IEEEtran.bst: No hyphenation pattern has been}%
\typeout{** loaded for the language `#1'. Using the pattern for}%
\typeout{** the default language instead.}%
\else
\language=\csname l@#1\endcsname
\fi
#2}}
\providecommand{\BIBdecl}{\relax}
\BIBdecl

\bibitem{carlozzi2016hdqlife}
N.~Carlozzi, S.~Schilling, J.-S. Lai, J.~Perlmutter, M.~Nance, J.~Waljee,
  J.~Miner, S.~Barton, S.~Goodnight, and P.~Dayalu, ``Hdqlife: the development
  of two new computer adaptive tests for use in huntington disease, speech
  difficulties, and swallowing difficulties,'' \emph{Quality of Life Research},
  vol.~25, no.~10, pp. 2417--2427, 2016.

\bibitem{mason2018predicting}
S.~L. Mason, R.~E. Daws, E.~Soreq, E.~B. Johnson, R.~I. Scahill, S.~J. Tabrizi,
  R.~A. Barker, and A.~Hampshire, ``Predicting clinical diagnosis in
  huntington's disease: An imaging polymarker,'' \emph{Annals of neurology},
  2018.

\bibitem{paulsen2010early}
J.~S. Paulsen, ``Early detection of huntington’s disease,'' \emph{Future
  neurology}, vol.~5, no.~1, pp. 85--104, 2010.

\bibitem{hinzen2017systematic}
W.~Hinzen, J.~Rossell{\'o}, C.~Morey, E.~Camara, C.~Garcia-Gorro, R.~Salvador,
  and R.~de~Diego-Balaguer, ``A systematic linguistic profile of spontaneous
  narrative speech in pre-symptomatic and early stage huntington's disease,''
  \emph{Cortex}, 2017.

\bibitem{vogel2012speech}
A.~P. Vogel, C.~Shirbin, A.~J. Churchyard, and J.~C. Stout, ``Speech acoustic
  markers of early stage and prodromal huntington's disease: A marker of
  disease onset?'' \emph{Neuropsychologia}, vol.~50, no.~14, pp. 3273--3278,
  2012.

\bibitem{HD_speech_characterization}
W.~P. Gordon and J.~Illes, ``Neurolinguistic characteristics of language
  production in huntington's disease: a preliminary report,'' \emph{Brain and
  Language}, vol.~31, no.~1, pp. 1--10, 1987.

\bibitem{dysarthria_1}
A.~B. Young, I.~Shoulson, J.~B. Penney, S.~Starosta-Rubinstein, F.~Gomez,
  H.~Travers, M.~A. Ramos-Arroyo, S.~R. Snodgrass, E.~Bonilla, H.~Moreno
  \emph{et~al.}, ``Huntington's disease in venezuela neurologic features and
  functional decline,'' \emph{Neurology}, vol.~36, no.~2, pp. 244--244, 1986.

\bibitem{dysarthria_2}
I.~Hertrich and H.~Ackermann, ``Acoustic analysis of speech timing in
  huntington′ s disease,'' \emph{Brain and language}, vol.~47, no.~2, pp.
  182--196, 1994.

\bibitem{kaploun2011acoustic}
L.~R. Kaploun, J.~H. Saxman, P.~Wasserman, and K.~Marder, ``Acoustic analysis
  of voice and speech characteristics in presymptomatic gene carriers of
  huntington's disease: biomarkers for preclinical sign onset?'' \emph{Journal
  of Medical Speech-Language Pathology}, vol.~19, no.~2, pp. 49--65, 2011.

\bibitem{dementia}
K.~C. Fraser, J.~A. Meltzer, N.~L. Graham, C.~Leonard, G.~Hirst, S.~E. Black,
  and E.~Rochon, ``Automated classification of primary progressive aphasia
  subtypes from narrative speech transcripts,'' \emph{cortex}, vol.~55, pp.
  43--60, 2014.

\bibitem{aphasia}
D.~Le, K.~Licata, C.~Persad, and E.~M. Provost, ``Automatic assessment of
  speech intelligibility for individuals with aphasia,'' \emph{IEEE/ACM
  Transactions on Audio, Speech, and Language Processing}, vol.~24, no.~11, pp.
  2187--2199, 2016.

\bibitem{toth2015automatic}
L.~T{\'o}th, G.~Gosztolya, V.~Vincze, I.~Hoffmann, G.~Szatl{\'o}czki,
  E.~Bir{\'o}, F.~Zsura, M.~P{\'a}k{\'a}ski, and J.~K{\'a}lm{\'a}n, ``Automatic
  detection of mild cognitive impairment from spontaneous speech using asr,''
  in \emph{Sixteenth Annual Conference of the International Speech
  Communication Association}, 2015.

\bibitem{fraser2013automatic}
K.~Fraser, F.~Rudzicz, N.~Graham, and E.~Rochon, ``Automatic speech recognition
  in the diagnosis of primary progressive aphasia,'' in \emph{Proceedings of
  the Fourth Workshop on Speech and Language Processing for Assistive
  Technologies}, 2013, pp. 47--54.

\bibitem{sadeghian2015using}
R.~Sadeghian, D.~J. Schaffer, and S.~A. Zahorian, ``Using automatic speech
  recognition to identify dementia in early stages,'' \emph{The Journal of the
  Acoustical Society of America}, vol. 138, no.~3, pp. 1782--1782, 2015.

\bibitem{zhou2016speech}
L.~Zhou, K.~C. Fraser, and F.~Rudzicz, ``Speech recognition in alzheimer's
  disease and in its assessment.'' in \emph{INTERSPEECH}, 2016, pp. 1948--1952.

\bibitem{mengistu2011comparing}
K.~Mengistu and F.~Rudzicz, ``Comparing humans and automatic speech recognition
  systems in recognizing dysarthric speech,'' \emph{Advances in Artificial
  Intelligence}, pp. 291--300, 2011.

\bibitem{peintner2008learning}
B.~Peintner, W.~Jarrold, D.~Vergyri, C.~Richey, M.~L.~G. Tempini, and J.~Ogar,
  ``Learning diagnostic models using speech and language measures,'' in
  \emph{Engineering in Medicine and Biology Society, 2008. EMBS 2008. 30th
  Annual International Conference of the IEEE}.\hskip 1em plus 0.5em minus
  0.4em\relax IEEE, 2008, pp. 4648--4651.

\bibitem{guerra2003modern}
E.~C. Guerra and D.~F. Lovey, ``A modern approach to dysarthria
  classification,'' in \emph{Engineering in Medicine and Biology Society, 2003.
  Proceedings of the 25th Annual International Conference of the IEEE},
  vol.~3.\hskip 1em plus 0.5em minus 0.4em\relax IEEE, 2003, pp. 2257--2260.

\bibitem{shoulson1989assessment}
I.~Shoulson, R.~Kurlan, A.~Rubin, D.~Goldblatt, J.~Behr, C.~Miller, J.~Kennedy,
  K.~A. Bamford, E.~D. Caine, D.~K. Kido \emph{et~al.}, ``Assessment of
  functional capacity in neurodegenerative movement disorders: Huntington’s
  disease as a prototype,'' \emph{Quantification of neurologic deficit. Boston:
  Butterworths}, pp. 271--283, 1989.

\bibitem{kieburtz2001unified}
K.~Kieburtz, J.~B. Penney, P.~Corno, N.~Ranen, I.~Shoulson, A.~Feigin,
  D.~Abwender, J.~T. Greenarnyre, D.~Higgins, F.~J. Marshall \emph{et~al.},
  ``Unified huntington’s disease rating scale: reliability and consistency,''
  \emph{Neurology}, vol.~11, no.~2, pp. 136--142, 2001.

\bibitem{Grandfather_p1}
J.~Reilly and J.~L. Fisher, ``Sherlock holmes and the strange case of the
  missing attribution: A historical note on “the grandfather passage”,''
  \emph{Journal of Speech, Language, and Hearing Research}, vol.~55, no.~1, pp.
  84--88, 2012.

\bibitem{duffy1995motor}
J.~Duffy, ``Motor speech disorders: Substrates, differential diagnosis, and
  management. st. louis, mo: Mosby-year book,'' 1995.

\bibitem{zraick2004reliability}
R.~I. Zraick, D.~J. Davenport, S.~D. Tabbal, T.~J. Hutton, G.~M. Hicks, and
  J.~H. Patterson, ``Reliability of speech intelligibility ratings using the
  unified huntington disease rating scale,'' \emph{Journal of Medical
  Speech-Language Pathology}, vol.~12, no.~1, pp. 31--41, 2004.

\bibitem{darley1975motor}
F.~L. Darley, A.~E. Aronson, and J.~R. Brown, \emph{Motor speech
  disorders}.\hskip 1em plus 0.5em minus 0.4em\relax Saunders, 1975.

\bibitem{patel2013caterpillar}
R.~Patel, K.~Connaghan, D.~Franco, E.~Edsall, D.~Forgit, L.~Olsen, L.~Ramage,
  E.~Tyler, and S.~Russell, ``“the caterpillar”: A novel reading passage
  for assessment of motor speech disorders,'' \emph{American Journal of
  Speech-Language Pathology}, vol.~22, no.~1, pp. 1--9, 2013.

\bibitem{macwhinney2000childes}
B.~MacWhinney, \emph{The CHILDES project: The database}.\hskip 1em plus 0.5em
  minus 0.4em\relax Psychology Press, 2000, vol.~2.

\bibitem{weide1998cmu}
R.~Weide, ``The cmu pronunciation dictionary, release 0.6,'' \emph{Carnegie
  Mellon University}, 1998.

\bibitem{povey2011kaldi}
D.~Povey, A.~Ghoshal, G.~Boulianne, L.~Burget, O.~Glembek, N.~Goel,
  M.~Hannemann, P.~Motlicek, Y.~Qian, P.~Schwarz \emph{et~al.}, ``The kaldi
  speech recognition toolkit,'' in \emph{IEEE 2011 workshop on automatic speech
  recognition and understanding}, no. EPFL-CONF-192584.\hskip 1em plus 0.5em
  minus 0.4em\relax IEEE Signal Processing Society, 2011.

\bibitem{roark2011spoken}
B.~Roark, M.~Mitchell, J.-P. Hosom, K.~Hollingshead, and J.~Kaye, ``Spoken
  language derived measures for detecting mild cognitive impairment,''
  \emph{IEEE transactions on audio, speech, and language processing}, vol.~19,
  no.~7, pp. 2081--2090, 2011.

\bibitem{GoP}
S.~M. Witt and S.~J. Young, ``Phone-level pronunciation scoring and assessment
  for interactive language learning,'' \emph{Speech communication}, vol.~30,
  no.~2, pp. 95--108, 2000.

\bibitem{abadi2016tensorflow}
M.~Abadi, P.~Barham, J.~Chen, Z.~Chen, A.~Davis, J.~Dean, M.~Devin,
  S.~Ghemawat, G.~Irving, M.~Isard \emph{et~al.}, ``Tensorflow: A system for
  large-scale machine learning.'' in \emph{OSDI}, vol.~16, 2016, pp. 265--283.

\bibitem{le2017automatic}
D.~Le, K.~Licata, and E.~M. Provost, ``Automatic paraphasia detection from
  aphasic speech: A preliminary study,'' \emph{Proc. Interspeech 2017}, pp.
  294--298, 2017.

\end{thebibliography}

\end{document}